\documentclass[12pt]{article}
\usepackage{physics}
\usepackage{epsf}
\usepackage{aas_macros}
\usepackage{amssymb}
\usepackage{comment}
\usepackage{amsmath}
\usepackage{braket}
\usepackage{mathrsfs}
\usepackage{mathtools}
\usepackage{array}
\usepackage{fancyhdr}
\usepackage[dvipdfmx]{graphicx}
\usepackage{color}
\usepackage{cite}
\usepackage{bm}
\usepackage{url}
\usepackage{fancyhdr}
\usepackage[colorlinks,citecolor=blue]{hyperref}
\usepackage[normalem]{ulem}

\usepackage[hang,small,bf]{caption}
\usepackage[subrefformat=parens]{subcaption}
\renewcommand{\thefootnote}{\#\arabic{footnote}}

\renewcommand{\thefootnote}{\fnsymbol{footnote}}
\def\thefootnote{\fnsymbol{footnote}}

\makeatletter

\@addtoreset{equation}{section}
\makeatother

\def\be{\begin{equation}}
\def\ee{\end{equation}}
\def\ben{\begin{eqnarray}}
\def\een{\end{eqnarray}}

\begin{document}


\begin{center}


{\Large \bf Initial clustering of primordial black holes: A general formulation for arbitrary local non-Gaussianity}

\vskip .5in
{\large
Hirotaka Ishijima, Teruaki Suyama, Ashu Kushwaha
}


{\em
Department of Physics, Institute of Science Tokyo, 2-12-1 Ookayama, Meguro-ku,
Tokyo 152-8551, Japan \\
}

\end{center}

\begin{abstract}

Initial spatial clustering of primordial black holes (PBHs) induced by local-type non-Gaussianity (LNG) can substantially modify cosmological constraints on PBH abundance.
Several inflationary scenarios that enhance curvature perturbations at small scales relevant to PBH formation predict LNG that is not necessarily perturbative. 
Therefore, it is crucial to establish a theoretical framework capable of investigating the initial clustering induced by arbitrary LNG.
Here, we present a general analytical formulation for the PBH two-point correlation function 
applicable to arbitrary LNGs in the alternative approach. 
Under the assumptions that PBHs form only at the large peaks of perturbations, and that large-scale modes weakly modulate the local variance of small-scale perturbations, we derive an analytic expression for the PBH bias parameter, 
directly connecting initial clustering to the primordial trispectrum in the collapsed limit. 
We demonstrate the versatility of our formula by computing the bias parameters in the ultra-slow-roll inflation, curvaton, and modulated reheating scenarios.
We also formally generalize the framework to broad power spectra to account for correlations across different PBH mass scales. 
Because our formulation does not rely on weak or perturbative non-Gaussianity assumptions,
our result provides a universal theoretical basis for evaluating initial clustering impact on PBH observables.

\end{abstract}

\renewcommand{\thepage}{\arabic{page}}
\setcounter{page}{1}
\renewcommand{\thefootnote}{\#\arabic{footnote}}
\setcounter{footnote}{0}

\newpage
\section{Introduction}
In the early universe, primordial black holes might have been formed in a non-astrophysical way \cite{Hawking-1971,Carr:1974nx}. 
Shortly after their initial proposal, PBHs were recognized as a viable dark matter (DM) candidate.
Over the decades, numerous observational constraints have been placed on their abundance across a wide range of masses (see Refs.~\cite{2018-Sasaki.etal-CQG,2020-Green.Kavanagh-JPhyG,Carr:2021bzv,2021-Carr.etal-Rept.Prog.Phys} for comprehensive reviews). 
Furthermore, since the LIGO's first historic detection of gravitational wave (GW) event, GW150914 \cite{LIGOScientific:2016dsl}, PBHs have garnered significant attention, as PBHs could serve the seeds of black hole mergers, which constitute the primary source of GW events \cite{Sasaki:2016jop,Clesse:2016vqa,Bird:2016dcv}. 

The standard scenario of PBH formation is the direct gravitational collapse of large-amplitude curvature perturbations during the radiation-dominated era.
In this scenario, the abundance, mass, spin, and spatial distribution of PBHs are determined by the statistics of curvature perturbations.
Therefore, understanding its effect on these PBH properties is crucial to estimate the cosmological observables related to PBHs.
Initial spatial clustering is particularly interesting because it can significantly affect PBH merger rates~\cite{Clesse:2016vqa,Ballesteros:2018swv,Young:2019gfc,Ding:2019tjk, Kasai:2026kpv}, microlensing signals\cite{Garcia-Bellido:2017xvr,Toshchenko:2025esh}, and the formation of the first structures \cite{Meszaros:1975ef,Carr:2018rid}.
These observables provide the basis for current constraints on the PBH abundance\cite{Carr:2021bzv}. 
However, existing constraints, including those from the EROS~\cite{EROS-2:2006ryy}, MACHO~\cite{Macho:2000nvd}, HSC~\cite{Niikura:2017zjd, Smyth:2019whb}, and OGLE~\cite{Niikura:2019kqi} microlensing surveys, 
typically assume that the effects of initial clustering are negligible at both early and late times 
(Ref.~\cite{2018-Sasaki.etal-CQG,2021-Carr.etal-Rept.Prog.Phys} for the recent reviews).
Since clustering can substantially weaken, or even remove, these constraints \cite{Garcia-Bellido:2017xvr, Belotsky:2018wph, Toshchenko:2025esh}, it may reopen the possibility that PBHs constitute all of DM.

Abundant PBH formation requires a significant enhancement of the primordial curvature perturbations 
at small scales, which necessitates inflationary scenarios beyond standard 
single-field slow-roll inflation \cite{Motohashi:2017kbs}.
Several such mechanisms have been proposed, including ultra slow-roll (USR) inflation and the curvaton scenario (see e.g., \cite{Pi:2024lsu}). 
In these scenarios, the curvature perturbations generally acquire local-type non-Gaussian statistics
which correlate small- and large-scale modes (see Section \ref{subsec:intuitive} for the definition of ``small'' and ``large'', and Section \ref{sec:benchmark} for the details on the correlation), leading to \textit{initial} PBH clustering\footnote{
Even if the statistics of curvature perturbation is completely Gaussian, PBHs spatially distribute as the Poisson distribution \cite{Ali-Haimoud:2018dau},
leading to the fluctuations of the number of PBHs $\sqrt{N}$ in a volume where the average number is $N$. 
In this work, \textit{initial clustering} refers to the deviation from the Poisson distribution.
}.
If this correlation is positive (negative), the amplitude of small-scale modes is amplified (suppressed) in regions where the underlying large-scale modes have a large (small) amplitude, thereby increasing (decreasing) the formation probability.
Consequently, PBHs become clustered at their formation time on the scale of the large-scale modes, which could substantially amplify clustering effects on cosmological observables. 
Therefore, it is important to accurately quantify the initial clustering of PBHs to assess clustering-induced modifications to constraints on the PBH abundance.

The PBH initial clustering in the presence of local-type non-Gaussianity (LNG) has intensively studied 
on either the small non-Gaussianity regime \cite{Tada:2015noa, Young:2015kda, Suyama:2019cst} 
or the specific type of extremely large non-Gaussianity \cite{Shinohara:2021psq, Kawasaki:2021zir, Kasai:2023ofh}. 
However, the types of LNG predicted by certain inflationary scenarios do not necessarily fall 
into these categories. 
Recent work by Iovino \textit{et.al.}~\cite{Crescimbeni:2025ywm} extended the analytical framework of the PBH two-point correlation function to a form applicable to arbitrary LNG. 
Assuming the amplitude of the large-scale curvature perturbation is sufficiently small, 
they formulated the linear bias parameter of PBHs, which scales the radiation fluctuations, 
by determining the peak statistics of the compaction function. 
By applying their framework to $\zeta = \chi + \frac{3}{5}f_{\mathrm{NL}}\chi^{2}$, 
they figured out the non-linear dependence of the PBH bias on the non-linearity parameter $f_{\mathrm{NL}}$.

Against this backdrop, in this work, we formulate the PBH correlation function without specifying the shape of LNG and its degree of non-Gaussianity by directly treating the statistics of the curvature perturbation. 
Similar to Ref.~\cite{Crescimbeni:2025ywm}, we assume that the small-scale perturbations are weakly modulated by the large-scale perturbations. 
However, we set the threshold onto the curvature perturbation and, 
to maintain the generality of our formulation, treat two Gaussian random field in LNG: 
one governing the small-scale perturbations, $\chi_\mathrm{S}$, and the other providing the large-scale modulation, $\psi_\mathrm{L}$ (see Section 3). 
This approach makes our derivation complementary to that of Ref.~\cite{Crescimbeni:2025ywm} and broadens the applicability of our formula—for example, 
to cases where different scalar fields generate $\chi_{\mathrm{S}}$ and $\psi_\mathrm{L}$. 
We subsequently apply this expression to investigate the resulting initial clustering for several widely studied benchmark examples of LNG, 
including the USR inflation, curvaton, and modulated reheating scenarios. 
Furthermore, when applied to the case $\zeta_\mathrm{S} = \chi_\mathrm{S} + \frac{3}{5}f_\mathrm{NL}\chi_\mathrm{S}\psi_\mathrm{L}$, 
our expression shows excellent agreement in the $f_\mathrm{NL}$-dependence with the results of Ref.~\cite{Crescimbeni:2025ywm}. 
These provide a theoretical foundation to rigorously investigate the effects of various types of LNG, via the initial clustering, on predicted cosmological observables and,
consequently, to revisit the observational constraints on PBHs.

This paper is organized as follows. 
In Section~\ref{sec: physical_picture}, we provide a qualitative illustration of the physical mechanism underlying PBH initial clustering.
Section~\ref{sec: derivation_of_correlator} starts with introduction of the definitions and notation used throughout this work.
Then, we derive the PBH two-point correlation function for an arbitrary LNG and explore the physical implications of our results by applying them to several specific models of non-Gaussian curvature perturbations. 
Finally, we summarize our findings and conclude in Section~\ref{sec: conlusion}.

\section{Qualitative description of initial clustering of PBHs}  \label{sec: physical_picture}

Before proceeding with the mathematical analysis, 
this section presents a qualitative explanation of the mechanism underlying 
the initial clustering of PBHs.

We consider a standard scenario where PBHs are formed in the radiation-dominated 
era from the direct gravitational collapse of large-amplitude primordial perturbations
originating from quantum fluctuations produced during the inflationary era.
In this scenario, the large overdensity is initially in the super-Hubble regime,
and PBH formation happens shortly after the overdensity reenters the Hubble horizon.

\subsection{Intuitive picture}
\label{subsec:intuitive}
In this subsection, we provide a model-independent explanation of how the initial clustering of PBHs is generated.

\begin{figure}[t]
    \centering
    \includegraphics[width=0.8\linewidth]{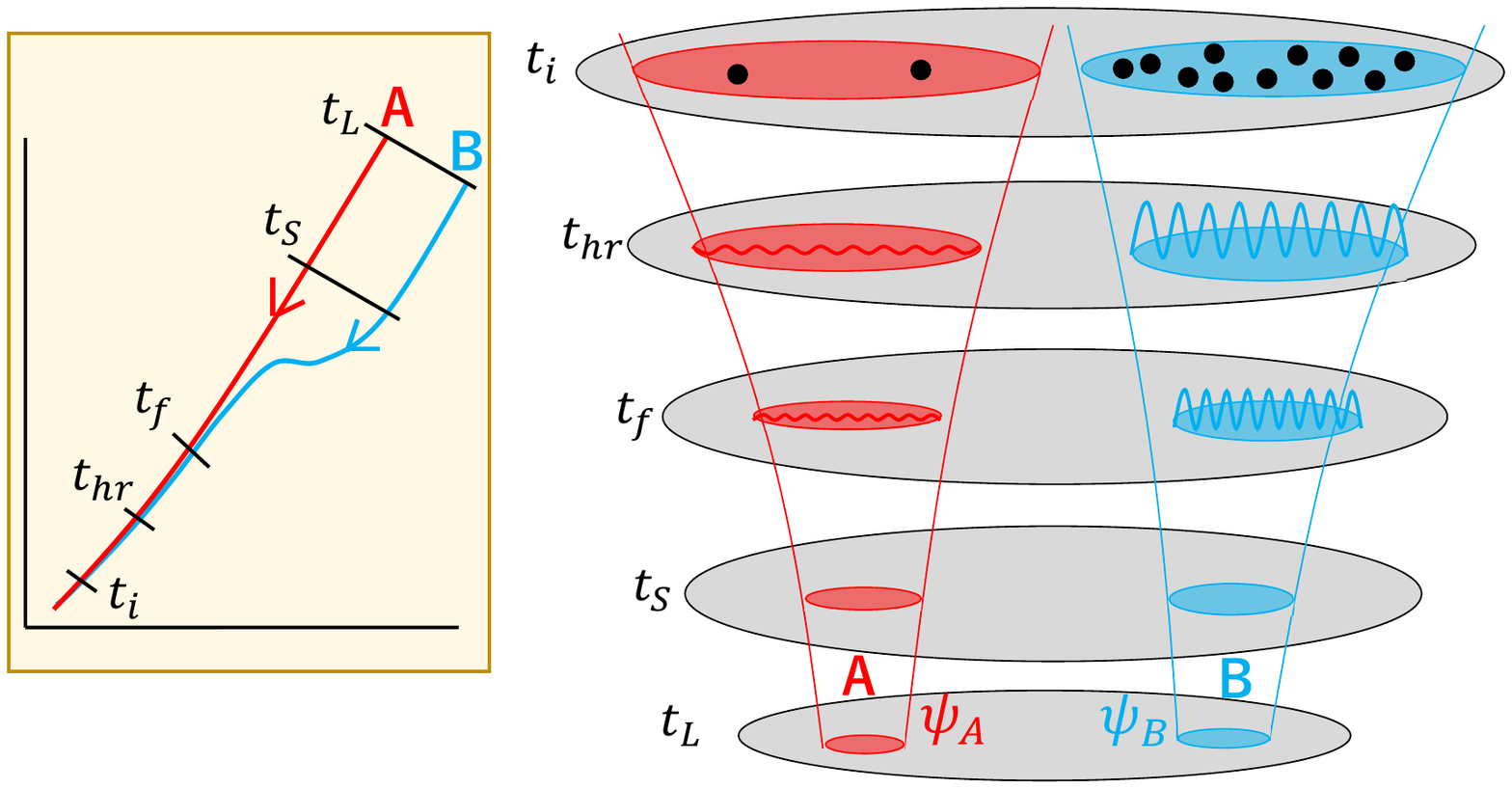}
    \caption{Schematic picture representing the mechanism to produce
    initial clustering of PBHs}
    \label{fig:Figure-separate}
\end{figure}

Fig.~\ref{fig:Figure-separate} schematically illustrates how the primordial perturbations lead to the clustering of PBHs.
In the most parts of this paper, to focus on the essential point and for the sake of simplicity,
we assume that primordial perturbations are significantly enhanced at scales peaked
around a specific scale
(which we refer to as the ``small scale'' for reasons that will become clear later) 
and that PBHs originate from such enhanced perturbations.
Extension to the case where the enhancement of the primordial perturbations extends over
a wide range of scales is discussed in \ref{extension-broad}.

We define $t_{\mathrm{S}}$ as the time that sets the initial conditions for the 
small-scale perturbations, which are in the super-Hubble regime. 
At time $t_{\mathrm{S}}$, we consider two non-overlapping comoving regions, A and B, 
both of which are much larger than the wavelength of the small-scale perturbations.
Thus, both regions contain many small-scale Hubble patches.
Hereafter, we call the size of A and B the ``large scale''.

Next, let us suppose that there exists a field $\psi$ that modulates at large scales and 
whose value affects the subsequent evolution of the universe.
Because the mechanism explained here operates as long as $\psi$ possesses 
large-scale perturbations and affects the cosmic expansion rate, 
we do not specify the precise nature of $\psi$ in this subsection.
We denote the values of $\psi$ in regions A and B at $t_{\mathrm{S}}$ as 
$\psi_\mathrm{A}$ and $\psi_\mathrm{B}$, respectively.
After the time $t_{\mathrm{S}}$, based on the separate universe approach \cite{Wands:2000dp}, 
each small-scale patch evolves as its own Friedmann-Lema\^{i}tre-Robaertson-Walker universe with an expansion history differing from one another. 
At a later time $t_\mathrm{f}$, while the small scale is still in the super-Hubble regime, 
we assume that all small-scale patches converge to the same evolution path 
as it is schematically represented by the left figure of Fig.~\ref{fig:Figure-separate}.
Consequently, after this time, all small-scale patches evolve in exactly 
the same manner, and the only quantity distinguishing individual patches is a time shift, which corresponds to the adiabatic curvature perturbation.
This perturbation remains constant in the super-Hubble regime \cite{Lyth:2004gb}.
According to the $\delta N$ formalism \cite{Starobinsky:1985ibc, Sasaki:1995aw, Sasaki:1998ug, Lyth:2004gb}, 
the final small-scale curvature perturbation $\zeta_{\mathrm{S}}$ is equal to 
the perturbation in the number of e-folds from $t_{\mathrm{S}}$ to $t_\mathrm{f}$.
Because regions A and B follow different trajectories due to the difference values
of $\psi$ in each patch (i.e., $\psi_\mathrm{A}$ and $\psi_\mathrm{B}$),
the variance of $\zeta_{\mathrm{S}}$ in region A will, in general, differ from that in region B, 
as schematically illustrated in the right figure of Fig.~\ref{fig:Figure-separate} 
(where region B exhibits a larger variance of $\zeta_{\mathrm{S}}$ than region A).
This implies that the variance of $\zeta_{\mathrm{S}}$ possesses a large-scale correlation,
signifying the presence of the curvature perturbation trispectrum 
in the collapsed limit \cite{Smith:2011if}.  
Eventually, the small-scale perturbations re-enter the Hubble horizon at 
time $t_{\rm hr}$, and any small-scale patches where the perturbation 
amplitude exceeds the threshold for PBH formation promptly collapse into PBHs.
As shown in Fig.~\ref{fig:Figure-separate}, perturbations in region B have 
a higher probability of exceeding the threshold than those in region A.
As a result, PBHs are produced more abundantly in region B than in region A, 
leading to the large-scale clustering of PBHs.

As is clear from the above discussion, 
the trispectrum in the collapsed limit acts as the source of PBH clustering.
The primary objective of this paper is to derive the mathematical 
expressions relating these two quantities, which we address in the subsequent sections.

\subsection{Benchmark scenarios generating local-type non-Gaussianity}
\label{sec:benchmark}
The discussion in the previous subsection is model-independent and valid whenever large-scale perturbation is correlated with small-scale perturbation responsible for the PBH formation.
In this subsection, we briefly provide a semi-qualitative and intuitive description of how this mechanism is realized in several benchmark scenarios that generate LNG, namely the USR inflation, curvaton, and modulated reheating scenarios.
A detailed analysis of the initial clustering in these models will be presented in Section~\ref{sec: example_LNG}, based on our expression derived in Section~\ref{sec: derivation_of_correlator}. \\

\textit{\underline{Standard single-field slow-roll scenario}}

We begin by discussing the simplest case which is the standard single-field slow-roll (SR) inflation.
Although LNG is absent and initial PBH clustering does not arise in this scenario,
we include this case as a first example to make it easier to understand a peculiar 
feature of USR inflation addressed below.
In the single-field SR scenario, inflaton $\phi$ rolls down the potential $V$ with
the background equation of motion given by ${\ddot \phi}+3H{\dot \phi}+V'(\phi)=0$.
The dot denotes the derivative with respect to the cosmic time $t$ and $H$ represents the Hubble parameter.
In the SR regime, the friction term is balanced with the potential term, reducing the original equation
to ${\dot \phi}\approx -\frac{V'}{3H}$ as well as the original Friedmann equation
to $H^2 \approx \frac{8\pi G}{3}V(\phi)$.
Combining these equations, ${\dot \phi}$ is uniquely determined by $\phi$, yielding a unique trajectory 
${\dot \phi}={\dot \phi}(\phi)$ in the field space spanned by $(\phi, {\dot \phi})$.
Even if ${\dot \phi}$ starts from a value different from the one given by the SR equations, 
the field quickly settles down to the trajectory.
Thus, SR is an attractor system.

\begin{figure}
    \centering
    \includegraphics[width=0.9\linewidth]{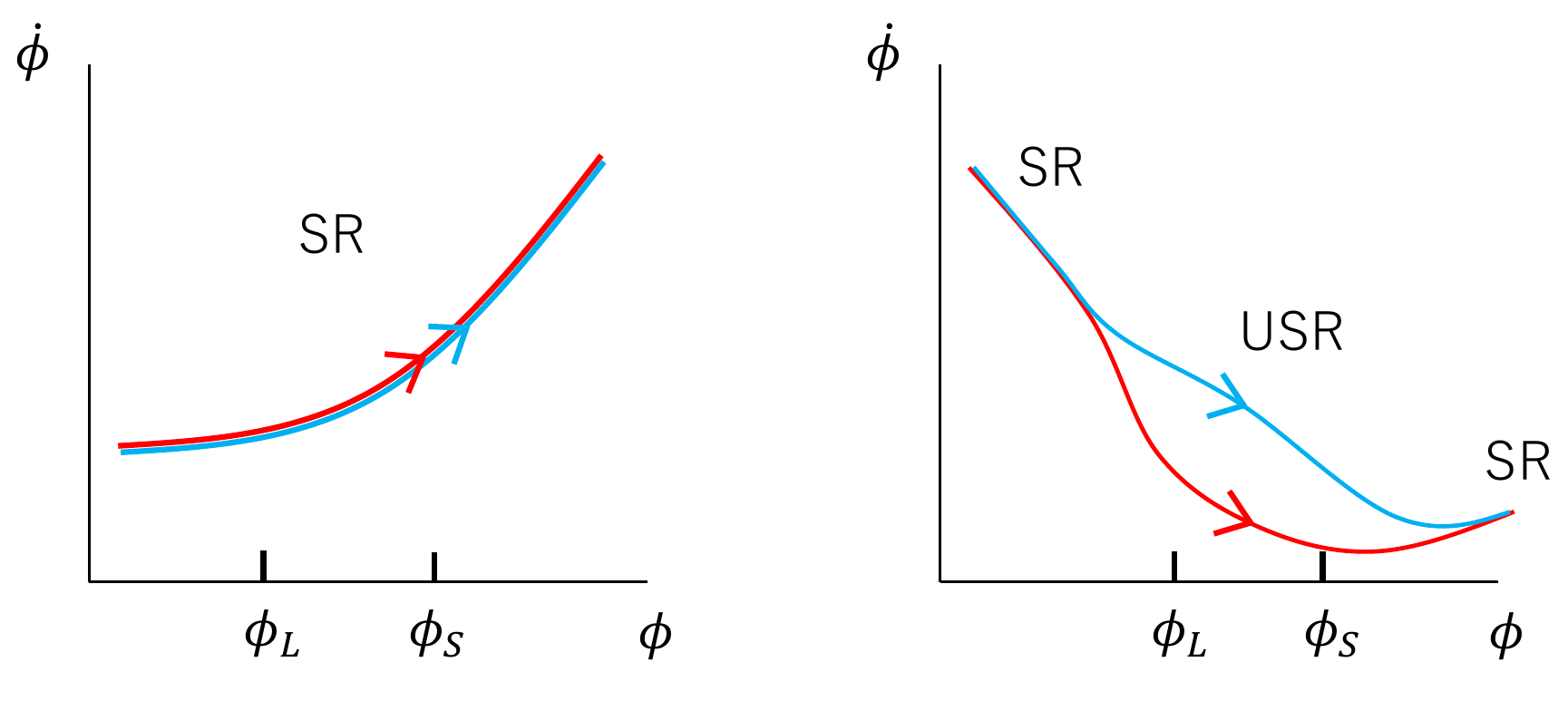}
    \caption{Left panel: Schematic picture showing trajectories in the $(\phi, {\dot \phi})$ space corresponding to two patches, A and B, during slow-roll inflation. 
    Right panel: Same as the left panel, but with an ultra-slow-roll phase between the slow-roll phases.}
    \label{fig:SR-USR}
\end{figure}

When large-scale classical perturbation $\delta \phi$ (originating from quantum fluctuations) 
is added to the background component, the field shifts along the background trajectory.
In terms of Fig.~\ref{fig:Figure-separate}, the region A and B follow the same evolution
(see also the left figure of Fig.~\ref{fig:SR-USR})
with only a time difference given by $\delta t=\frac{\delta \phi_\mathrm{A}-\delta \phi_\mathrm{B}}{\dot \phi}$.
Thus, the small-scale curvature perturbation $\zeta_{\mathrm{S}}$ has no correlation with the large-scale
curvature perturbation,
yielding vanishing bispectrum in the squeezed limit (and trispectrum in the collapsed limit) \footnote{
In single-field SR inflation, it is known that
the non-linearity parameter $f_{\rm NL}$ of the LNG, which characterizes
the magnitude of the bispectrum in the squeezed limit, is given by
$f_{\rm NL}=\frac{5}{12} (1-n_s)$ \cite{Maldacena:2002vr}, 
where $n_s$ is the spectral index of the power spectrum
of curvature perturbations.
Thus, if $n_s \neq 1$, $f_{\rm NL}$ is non-zero. 
However, this $f_{\rm NL}$ is a gauge artifact and 
does not represent a physical correlation between small-scale and large-scale perturbations \cite{Tanaka:2011aj, Pajer:2013ana}.
Similarly, a non-zero $\tau_{\rm NL} =\frac{1}{4} {(1-n_s)}^2 (=\frac{36}{25}f_{\rm NL}^2)$ is also a gauge artifact \cite{Suyama:2020akr}.
Whenever the non-linearity parameters are presented in this paper, we implicitly assume that
the part corresponding to the gauge artifact has been already subtracted.
}.\\

\textit{\underline{Ultra slow-roll scenario}}

In the USR scenario \cite{Tsamis:2003px, Kinney:2005vj}, inflaton field $\phi$ 
has a potential which is sufficiently flat in some field range. 
When the field is in that range, the background equation of motion of $\phi$ becomes
${\ddot \phi}+3H {\dot \phi}\approx 0$.
To be consistent with the observation, such a USR phase should lie on enough small scales while remaining the prediction on the CMB scales as the prediction from the standard SR inflation.
In the USR phase, approximating $H$ as constant, the solution of the equation of motion is given by
$\phi (t)=\phi_0+\frac{v_0}{3H} (1- e^{-3Ht})$,
where $\phi_0$ and $v_0$ are values of $\phi$ and ${\dot \phi}$ at $t=0$.
Thus, the inflaton's {\it velocity} becomes ${\dot \phi}=v_0 e^{-3Ht}$ and
decays exponentially to zero with time (hence {\it ultra}-slow-roll).
Contrary to the standard SR inflation, the velocity does not approach a unique
value determined by the field value, meaning that the USR is not an attractor system 
as it is schematically represented in the right figure of Fig.~\ref{fig:SR-USR}.
As a result, the patches A and B at the time $t_{\mathrm{S}}$ are not in the same state due to the existence
of the large-scale perturbation of $\phi$.

From the above solution with the initial conditions $\phi (t_\mathrm{S})=\phi_{\mathrm{S}}$ and ${\dot \phi}_\mathrm{S} (t_\mathrm{S})=v_\mathrm{S}$, 
the number of e-folds required for $\phi$ to reach $\phi_\mathrm{f}$ where the USR phase ends is given by
\begin{equation}
N_\mathrm{S}=-\frac{1}{3} \ln \left( 1-\frac{3H}{v_\mathrm{S}} (\phi_\mathrm{f}-\phi_{\mathrm{S}}) \right).
\end{equation}
If the large-scale fluctuation $\delta \phi_{\mathrm{L}}$, which is generated earlier than $t_{\mathrm{S}}$, 
is added to $\phi_{\mathrm{L}}$ in the large-scale patch (say patch A in Fig.~\ref{fig:Figure-separate}),
$\phi_{\mathrm{S}}$ in the same patch receives a shift by $\delta \phi_{\mathrm{L}}$ (in the language of the previous subsection, $\delta \phi_{\mathrm{L}}$ corresponds to $\psi$).
As a result, the small-scale curvature perturbation in this patch becomes \cite{Namjoo:2012aa}
\begin{equation} \label{USR-zetas}
\zeta_{\mathrm{S}}=-\frac{1}{3} \ln \left( 1+\frac{3H}{v_\mathrm{f}} (\delta \phi_{\mathrm{S}}+\delta \phi_{\mathrm{L}}) \right)+
\frac{1}{3} \ln \left( 1+\frac{3H}{v_\mathrm{f}} \delta \phi_{\mathrm{L}} \right),
\end{equation}
where $v_\mathrm{f}=v_{\mathrm{S}}-3H(\phi_{\mathrm{f}}-\phi_{\mathrm{S}})$ is the value of ${\dot \phi}$ at $t_\mathrm{f}$ 
and the second term has been added to ensure that $\zeta_{\mathrm{S}}$
vanishes when $\delta \phi_{\mathrm{S}} =0$.

Because of the logarithmic function, $\delta \phi_{\mathrm{S}}$ is coupled to large-scale perturbation $\delta \phi_{\mathrm{L}}$.
To see this more explicitly, let us expand Eq.~(\ref{USR-zetas}) up to quadratic order in fluctuations as
\begin{equation}
\zeta_{\mathrm{S}} \approx \chi_{\mathrm{S}} +\frac{3}{2} \left( \chi_{\mathrm{S}}^2 +2\chi_{\mathrm{S}} \psi_{\mathrm{L}} \right),
\end{equation}
where we have introduced $\chi_{\mathrm{S}} =-\frac{H}{v_\mathrm{f}}\delta \phi_{\mathrm{S}}, ~\psi_{\mathrm{L}}=-\frac{H}{v_\mathrm{f}}\delta \phi_{\mathrm{L}}$.
Obviously, $\psi_\mathrm{L}$ enters $\zeta_{\mathrm{S}}$ through the coupling with $\chi_{\mathrm{S}}$.
Comparing this expression with the standard expression with the local-type non-linearity parameter $f_{\rm NL}$ given by $\zeta_{\mathrm{S}} =\chi_{\mathrm{S}}+\frac{3}{5}f_{\rm NL} (\chi_{\mathrm{S}}^2+2\chi_{\mathrm{S}} \psi_\mathrm{L})$,
it is immediate to find that $f_{\rm NL}=\frac{5}{2}$ in the present case \cite{Namjoo:2012aa, Martin:2012pe}.
In drawing this conclusion, it is assumed that the phase of USR abruptly terminates and
the universe quickly recovers the SR dynamics, which may be artificial. 
If the transition from USR to SR is made smoother, the resulting $f_{\rm NL}$ is suppressed
compared to the abrupt case \cite{Cai:2018dkf, Passaglia:2018ixg, Suyama:2021adn, Pi:2022ysn}.\\

\textit{\underline{Curvaton scenario}}

In the single-field inflation scenario, inflaton not only drives inflation but also
generates primordial curvature perturbation from its quantum fluctuations.
However, this may be not the case. 
It is possible that inflaton only drove inflation and the primordial curvature perturbations were
dominantly created by another field.
A curvaton scenario is one representative scenario realizing this possibility \cite{Lyth:2001nq, Enqvist:2001zp, Moroi:2001ct, Sasaki:2006kq}.
In this scenario, a light scalar field $\varphi$ called \textit{curvaton} is introduced in addition to inflaton field.
The curvaton acquires classical perturbations in the standard way during inflation, but 
little contributes to the total energy density during inflation.
The curvaton's field value remains frozen during both inflation and early stage of the subsequent
radiation-dominated era.
When the Hubble parameter becomes comparable to the curvaton mass, the curvaton
starts to oscillate after which the curvaton acts as a matter-like component and
the curvaton energy density grows relative to radiation energy density.
Eventually, the curvaton decays into lighter particles which quickly turn into thermal radiation.
If curvaton becomes a dominant component before the decay,
the curvature perturbation is generated by the curvaton perturbation 
when the curvaton-dominated era starts.
If curvaton remains a subdominant component throughout,
the curvature perturbation is generated when the curvaton decays. 
Adopting the sudden-decay approximation,
the final curvature perturbation $\zeta$ generated by the curvaton is obtained as a solution to an algebraic equation given by \cite{Sasaki:2006kq}
\begin{equation}
e^{4\zeta}-\Omega_{\varphi,\mathrm{d}} \frac{\rho_{\varphi}}{\bar{\rho}_\varphi} e^\zeta+
\Omega_{\varphi,\mathrm{d}}-1=0,
\end{equation}
where $\rho_\varphi$ is the curvaton energy density,
$\Omega_{\varphi,\mathrm{d}}=\frac{\bar{\rho}_\varphi}{\bar{\rho}_\mathrm{r}+\bar{\rho}_\varphi}\big|_\mathrm{d}$ is the
fraction of curvaton energy density at the time of the curvaton decay, 
and $\rho_{\mathrm{r}}$ is the radiation energy density.
The bar indicates that the corresponding quantity is evaluated at the background.
In what follows, we consider a limiting case $\Omega_{\varphi,d}=1$ 
where the curvaton once dominates the universe before the decay.
In this case, the above equation can be easily solved and the solution is given by
\begin{equation}
\zeta=\frac{1}{3} \ln \frac{\rho_{\varphi}}{\bar{\rho}_\varphi}.
\end{equation}
Treating the curvaton as a free field, we have $\rho_\varphi =\frac{m_\varphi^2}{2}\varphi^2$ ($m_\varphi$ is curvaton's mass).
Writing the curvaton field as $\varphi={\bar \varphi}+\delta \varphi_{\mathrm{S}}+\delta \varphi_{\mathrm{L}}$,
the small-scale curvature perturbation becomes
\begin{equation}
\zeta_{\mathrm{S}}=\frac{1}{3}\ln {\left( 1+\frac{\delta \varphi_{\mathrm{S}}}{\bar \varphi}+\frac{\delta \varphi_{\mathrm{L}}}{\bar \varphi} \right)}^2-
\frac{1}{3}\ln {\left( 1+\frac{\delta \varphi_{\mathrm{L}}}{\bar \varphi} \right)}^2.
\end{equation}
This is similar to $\zeta_{\mathrm{S}}$ in the USR scenario (see Eq.~(\ref{USR-zetas})),
and hence $\zeta_{\mathrm{S}}$ modulates on large scale due to modulation of $\delta \varphi_{\mathrm{L}}$.
Making the similar expansion as in the USR case, the small-scale curvature perturbation 
up to second order in fluctuations becomes
\begin{equation}
\zeta_{\mathrm{S}} =\chi_{\mathrm{S}}-\frac{3}{4} (\chi_{\mathrm{S}}^2+2\chi_{\mathrm{S}} \psi_{\mathrm{L}}),
\end{equation}
where $\chi_{\mathrm{S}}=\frac{2}{3} \frac{\delta \varphi_{\mathrm{S}}}{\bar \varphi},~\psi_{\mathrm{L}}=\frac{2}{3} \frac{\delta \varphi_{\mathrm{L}}}{\bar \varphi}$.
Thus, $f_{\rm NL}=-\frac{5}{4}$ in the present case \cite{Lyth:2002my}.

Although the curvaton scenario was originally proposed to explain the primordial perturbations on
CMB scales, we do not require the perturbations on such scales originate from curvaton even when
PBHs are assumed to originate from curvaton perturbations.
This is possible if the curvaton perturbations are largely enhanced on the PBH scale
and such cases have been investigated in the literature \cite{Kawasaki:2012wr,Kohri:2012yw, Ando:2017veq,Pi:2021dft,Ferrante:2023bgz,Gow:2023zzp,Chen:2024pge, Kuroda:2025coa}.\\

\textit{\underline{Modulated reheating scenario}}

Similar to the curvaton scenario, the modulated reheating scenario also invokes an additional field other than the inflaton to generate primordial curvature perturbations.
In this scenario, the inflaton decay rate is modulated by one or more spectator fields \cite{Dvali:2003em, Kofman:2003nx, Zaldarriaga:2003my,Suyama:2007bg, You:2024hit}.
Assuming the universe expands like a matter-dominated universe during the oscillating phase of the inflaton, which is the case if the inflaton potential around its minimum is well approximated by a quadratic form,
the total number of e-folds from the end of inflation to a certain time in the radiation-dominated era depends on the decay rate $\Gamma$.
To see this, let us adopt a sudden-decay approximation where the inflaton instantly decays into radiation at the time when $H=\Gamma$.
Under this approximation, the inflaton energy density decreases like matter until the time of decay.
Thus, the number of e-folds $N_1$ between the onset of the inflaton's oscillation and the decay is given by
\begin{equation}
N_1=-\frac{1}{3} \ln \left( \frac{3\Gamma^2}{8\pi G \rho_{\rm ini}}\right),
\end{equation}
where $\rho_{\rm ini}$ is the initial energy density of the inflaton.
After the decay, the universe becomes radiation-dominated.
Thus, the number of e-folds $N_2$ from the decay until the radiation energy density reaches a certain value $\rho_\mathrm{f}$ is given by
\begin{equation}
N_2=-\frac{1}{4} \ln \left( \frac{8\pi G\rho_\mathrm{f}}{3\Gamma^2} \right).
\end{equation}
Then, the total number of e-folds $N$ is $N_1+N_2$ and depends on $\Gamma$ as 
\begin{equation}
N=-\frac{1}{6} \ln \Gamma +\cdots
\end{equation}
where $\cdots$ denotes terms independent of $\Gamma$.
If $\Gamma$ is a function of a light field $\varphi$ that acquires classical perturbations during inflation, the small-scale curvature perturbation becomes
\begin{equation}
\label{zetas-modulated}
\zeta_{\mathrm{S}} =-\frac{1}{6} \ln \Gamma ({\bar \varphi}+\delta \varphi_{\mathrm{S}}+\delta \varphi_{\mathrm{L}})+
\frac{1}{6} \ln \Gamma ({\bar \varphi}+\delta \varphi_{\mathrm{L}}).
\end{equation}
Similar to the USR scenario, the logarithmic function induces coupling between small-scale and large-scale perturbations. 
Furthermore, the non-linear dependence of $\Gamma$ on $\varphi$, which is the case for standard types of interactions, introduces additional coupling 
between small-scale and large-scale perturbations.
Expanding Eq.~(\ref{zetas-modulated}) up to quadratic order, we obtain
\begin{equation}
\zeta_{\mathrm{S}}=\chi_{\mathrm{S}}+3 \left( 1-\frac{\Gamma \Gamma''}{(\Gamma')^{2}} \right) (\chi_{\mathrm{S}}^2
+2\chi_{\mathrm{S}} \psi_\mathrm{L}),
\end{equation}
where $\chi_{\mathrm{S}}=-\frac{1}{6} \frac{\Gamma'}{\Gamma} \delta \varphi_\mathrm{S},~\psi_\mathrm{L}=-\frac{1}{6} \frac{\Gamma'}{\Gamma} \delta \varphi_\mathrm{L}$ and $\Gamma$
and its derivatives with respect to $\varphi$ are evaluated for the background value.
Thus, $f_{\rm NL}=5\left( 1-\frac{\Gamma \Gamma''}{(\Gamma')^{2}} \right)$ in the present case  \cite{Zaldarriaga:2003my, Suyama:2007bg}.

\section{Formulation of PBH correlation function}  
In this section, we derive a generic formula of the two-point correlation function of the initial PBH spatial distribution induced from an arbitrary LNG.

\label{sec: derivation_of_correlator}
\subsection{Definitions and assumptions}  \label{subsec: definitions}
In this subsection, we provide the relevant definitions and assumptions
which are used in this work. 
We are interested in the initial clustering of PBHs generated from LNGs of curvature perturbations, 
in which comoving curvature perturbation $\zeta$ at a position $\vec{x}$ is given as the non-linear function of the uncorrelated Gaussian variables at the same position 
$\vec{x}$:
\begin{equation} \label{def: general_LNG}
    \zeta(\vec{x}) = f( \chi(\vec{x}), \psi(\vec{x}), \cdots ).  
\end{equation}

The Gaussian variables $\{\chi, \psi, \cdots\}$ can physically be a fluctuation of a scalar field generated during the inflationary era (for example, see \cite{Sasaki:2006kq}).
However, in the present and the next sections, 
we do not need to specify their physical origin and simply treat them as Gaussian random fields.
This makes sure that our derivation of the two-point correlation function of PBHs can be applied to any LNG, 
as we describe in Sec.~\ref{sec: example_LNG}

To simplify the analysis, we assume that only two Gaussian fields, 
$\chi$ and $\psi$, are responsible for the curvature perturbation on small scales.
Specifically, $\chi$, which fluctuates on small scales, acts as the primary source of the curvature perturbation relevant to PBH formation, 
while $\psi$ provides large-scale modulation.
This is the minimal setup capable of capturing the initial clustering of PBHs, as illustrated in the previous section.
Hereafter, we denote $\chi$ and $\psi$ as $\chi_{\mathrm{S}}$ and $\psi_{\mathrm{L}}$, respectively.
Another important assumption made in this paper is that $\psi_{\mathrm{L}}$ 
only weakly modulates the small-scale curvature perturbation, allowing us to treat the 
dependence on $\psi_{\mathrm{L}}$ to the lowest order in its amplitude.
On the other hand, we do not make any assumptions about the dependence on $\chi_{\mathrm{S}}$,
thereby allowing for cases where the dependence is highly non-linear.

Generalizing the discussions in \ref{sec:benchmark}, $\zeta_{\mathrm{S}}$ is given by
\begin{equation}  \label{def: F}
    \zeta_{\mathrm{S}}\left(\vec{x}\right) = f(\chi_{\textrm{S}}(\vec{x}),\psi_{\textrm{L}}(\vec{x}))-f(0,\psi_{\textrm{L}}(\vec{x})) \equiv F (\chi_{\mathrm{S}}\left(\vec{x}\right), \psi_{\mathrm{L}}\left(\vec{x}\right) ).
\end{equation}
In the following analysis, we adopt a simplified criterion for PBH formation, 
assuming that a PBH forms if $\zeta_{\mathrm{S}}$ exceeds a threshold value $\zeta_{\mathrm{S, th}}$. 
This criterion has been extensively refined over the decades.
For instance, the formation threshold is known to depend on the perturbation profile \cite{Polnarev:2006aa, Nakama:2013ica, Musco:2018rwt, Escriva:2020tak}.
Yet, we maintain this simpler approach here. 
Furthermore, other local quantities 
such as the density perturbation on comoving slices and the compaction function are often used to quantify PBH formation \cite{Shibata:1999zs, Harada:2013epa, Harada:2015yda, Musco:2018rwt}. 
Incorporating these features may yield additional corrections to the formula we present later.

For later convenience, we also introduce a function $G$  
\begin{equation} \label{def: G}
    \chi_{\mathrm{S}} \equiv G (\zeta_{\mathrm{S}}, \psi_{\mathrm{L}}),  
\end{equation}
defined as the inverse of $F$ in Eq.~\eqref{def: F}\footnote{
In general, it is not trivial that a real-valued $G$ exists. For example, in the case of the power-series LNG up to the quadratic order, $F$ is given as Eq.~\eqref{eq: quad_F}.
The condition for $G$ to be real is equivalent to the existence of real solutions to the quadratic equation for $\chi_{\mathrm{S}}$, such that
\[
    \zeta_{\mathrm{S}} > -\frac{5}{12f_{\mathrm{NL}}}\left(\frac{6}{5}f_{\mathrm{NL}}\psi_{\mathrm{L}}+1\right)^{2}
\]
is satisfied. However, this is not an issue for our current work. 
Since we generally focus on the positive and large-amplitude regime of $\zeta_{\textrm{S}}$ in the discussion of PBH formation, this condition is naturally satisfied.
}.
In general, the function $G$ is not necessarily a single-valued function (see Sec.~\ref{sec: example_LNG}).

To quantify the PBH clustering, we focus on the two-point correlation function of the PBH spatial distribution, defined as \cite{Suyama:2019cst}
\begin{equation}  \label{def: pbh_correlator}
    \xi_{\mathrm{PBH}}(\vec{x}_{1}, \vec{x}_{2}) \equiv \frac{P_{2}(\vec{x}_{1}, \vec{x}_{2})}{P_{1}(\vec{x}_{1})P_{1}(\vec{x}_{2})}-1,
\end{equation}
where $P_{1}(\vec{x})$ is the probability that a PBH is formed in a region centered at $\vec{x}$, and $P_{2}(\vec{x}_{1}, \vec{x}_{2})$ is the joint probability that two PBHs are formed simultaneously in two regions centered at $\vec{x}_{1}$ and $\vec{x}_{2}$.
In the following subsections, we derive general expressions of $P_1$ and $P_2$, respectively.

\subsection{Formation probability in one Hubble patch}
In this subsection, we fix the position $\vec{x}$ and $\psi_{\mathrm{L}}$ and formulate the formation probability in one Hubble patch centered at $\vec{x}$, denoted by $P_{\mathrm{form}}(\psi_{\mathrm{L}}(\vec{x}))$ (or equivalently 
$P_{\mathrm{form}}(\psi_{\mathrm{L}})$).
Once $P_{\mathrm{form}}(\psi_{\mathrm{L}}(\vec{x}))$ is obtained, $P_1$ and $P_2$ can be computed
by taking ensemble average of quantities consisting of $P_{\mathrm{form}}(\psi_{\mathrm{L}})$ over $\psi_{\mathrm{L}}$.

\begin{figure}
  \centering
  \begin{subfigure}[b]{0.45\textwidth}
    \centering
    \includegraphics[width=\textwidth]{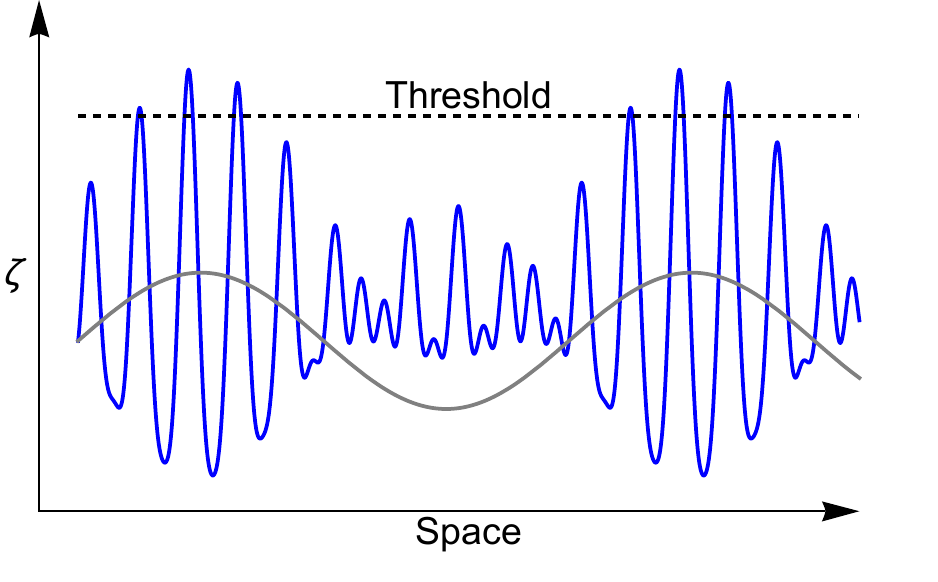}
    \caption{}
    \label{fig:sub-left}
  \end{subfigure}
  \hfill 
  \begin{subfigure}[b]{0.45\textwidth}
    \centering
    \includegraphics[width=\textwidth]{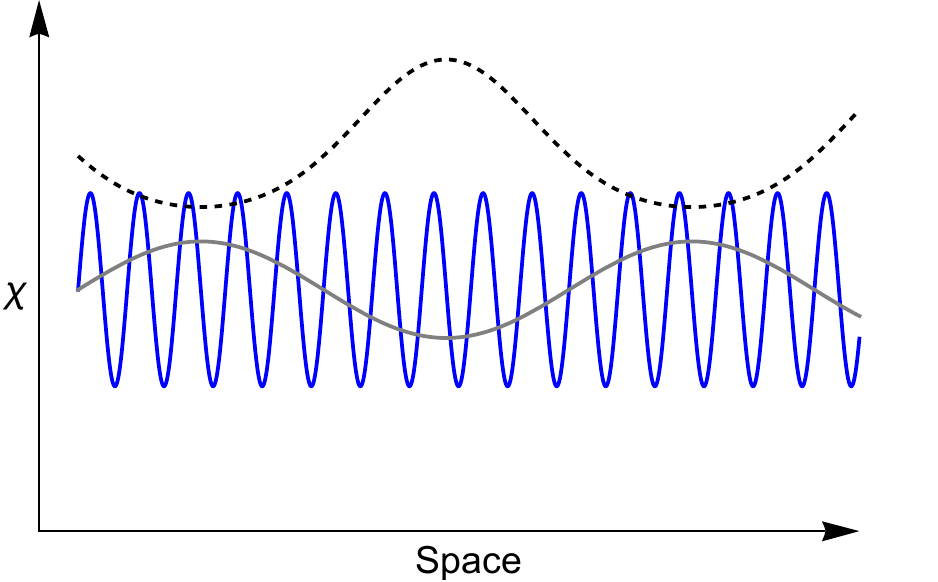}
    \caption{}
    \label{fig:sub-right}
  \end{subfigure}
  
  \caption{The difference of the PBH formation condition for (a) the curvature perturbation $\zeta$ and (b) Gaussian variable $\chi$. The blue (gray) line denotes the fluctuation at small (large) scales and the black dashed line is threshold in each figure. The amplification for the curvature perturbation is expressed as the modification for the threshold for the Gaussian variable.}
  \label{fig: schematic_fluctuation}
\end{figure}

Since PBHs are assumed to form in regions with $\zeta_{\mathrm{S}}$ exceeding the threshold $\zeta_{\mathrm{S, th}}$, 
$P_{\mathrm{form}}(\psi_{\mathrm{L}})$ is given by
\begin{equation} \label{eq: P1_chi}
    P_{\mathrm{form}}(\psi_{\mathrm{L}}) = \int_{F(\chi_{\mathrm{S}}, \psi_{\mathrm{L}})\geq\zeta_{\mathrm{S, th}}}d\chi_{\mathrm{S}}\frac{1}{\sqrt{2\pi}\sigma_{\mathrm{S}}}\exp\left[-\frac{\chi^{2}_{\mathrm{S}}}{2\sigma_{\mathrm{S}}^{2}}\right], 
\end{equation}
where $\sigma_{\mathrm{S}}$ is the variance of $\chi_{\mathrm{S}}$.
This shows that, in terms of the Gaussian variable $\chi_{\mathrm{S}}$, the large-scale mode 
$\psi_{\mathrm{L}}$ modulates the threshold value for the PBH formation depending on the position, 
while leaving the amplitude of the small-scale mode $\chi_{\mathrm{S}}$ unchanged, as illustrated in Fig.~\ref{fig: schematic_fluctuation}.
In terms of Eq.~\eqref{def: G}, all upper and lower limit of the solution of this inequality are given as
\begin{equation}  \label{def: chi_th}
    \chi_{\mathrm{S, th}} \equiv G(\zeta_{\mathrm{S, th}}, \psi_{\mathrm{L}}).
\end{equation}
Here, since $G$ is not necessarily a one-to-one function, we generally have multiple $\chi_{\mathrm{S}, \mathrm{th}}$ depending on the concrete form of $F$.
This corresponds to the fact that the solution of the inequality, in general, can be the multiple disconnected regions in the Gaussian variable $\chi_{\mathrm{S}}$.
Although this appears to complicate the analysis, this difficulty is safely bypassed in our case
by noting that the PBH formation is typically analyzed under the high-peak condition, $\zeta_{\mathrm{S, th}} \gg \sqrt{\langle\zeta_{\mathrm{local}}^{2}\rangle}$, which implies that the formation probability is dominated by the integration over the tail of the probability density function (PDF) of $\zeta_{\mathrm{S}}$.
Accordingly, the relevant intervals in the Gaussian variable $\chi_{\mathrm{S}}$ must also correspond to the tail of its Gaussian PDF.
Therefore, the dominant contribution to $P_{\mathrm{form}}$ arises from the region whose lower boundary is the minimum solution for $\chi_{\mathrm{S},\mathrm{th}}$\footnote{
For the case that multiple integration intervals comparably contribute to $P_{\mathrm{form}}$, it is straightforward to generalize our analysis.
}.
Furthermore, we extend the upper limit of the integration interval to infinity, as the integral over the PDF tail is overwhelmingly dominated by the contribution near the lower limit. 
Here and hereafter, $\chi_{\mathrm{S},\mathrm{th}}$ denotes the minimum solution of Eq.~\eqref{def: chi_th} (and the same prescription is adopted for $G(\zeta_{\mathrm{S, th}},\phi_{\mathrm{L}})$).

Under these assumptions, $P_{\mathrm{form}}$ can be approximated as
\begin{equation} \label{eq: Plocal_hpl}
    P_{\mathrm{form}}( \psi_{\mathrm{L}}) = \int_{\nu(\psi_{\mathrm{L}})}^{\infty}d\bar{\chi}_{\mathrm{S}}\frac{1}{\sqrt{2\pi}}\exp\left[-\frac{\bar{\chi}_{\mathrm{S}}^{2}}{2}\right] = \frac{1}{\sqrt{2\pi}\nu}e^{-\nu^{2}/2} \left( 1+\mathcal{O}(\nu^{-2})\right),  
\end{equation}
where we have defined the high-peak parameter 
$\nu (\psi_{\mathrm{L}},
\vec {x})\equiv\chi_{\mathrm{S},\mathrm{th}}(\vec{x})/\sigma_{\mathrm{S}}$. 
In the second equality, we have imposed the high-peak limit, 
$\nu\gg1$ and kept only the leading order of the integral 
in terms of the inverse of $\nu$.

As we have mentioned in the previous subsection, $\psi_{\mathrm{L}}$ is assumed to only weakly modulate $\zeta_{\mathrm{S}}$ on large scales.
Based on this assumption, we expand the threshold $\nu$ in terms of 
$\psi_{\mathrm{L}}$ as
\begin{equation}  \label{eq: expanded_nu}
    \nu(\psi_{\mathrm{L}}) = \frac{G(\zeta_{\mathrm{S, th}}, \psi_{\mathrm{L}})}{\sigma_{\mathrm{S}}} = \frac{1}{\sigma_{\mathrm{S}}}
    \sum_{n=0}^{\infty}\frac{1}{n!}G^{(n)}
    \psi_{\mathrm{L}}^{n},
\end{equation}
where
\begin{equation}  \label{def: g_bar}
    G^{(n)} \equiv \frac{\partial^{n}G\left(\zeta_{\mathrm{S, th}},\psi_{\mathrm{L}}\right)}{\partial\psi_{\mathrm{L}}^{n}}\Bigg|_{\psi_{\mathrm{L}}=0}.
\end{equation}
Truncating Eq.~\eqref{eq: expanded_nu} up to quadratic order in $\psi_{\mathrm{L}}$ and substituting it into Eq.~\eqref{eq: Plocal_hpl}, 
the formation probability in a single Hubble patch with the underlying long-wavelength mode $\psi_{\mathrm{L}}$ is expressed as
\begin{equation}  \label{eq: resultant_Pform}
    P_{\mathrm{form}}(\psi_{\mathrm{L}}) \simeq \frac{1}{\sqrt{2\pi}\bar{\nu}} e^{-\bar{\nu}^{2}/2} \left(1 + A_{1} \psi_{\mathrm{L}} + A_{2} \psi_{\mathrm{L}}^{2}\right),
\end{equation}
where
    \begin{align} \label{def: A}
        \bar{\nu} &\equiv \nu (\psi_{\mathrm{L}}=0), \\
        A_{1} &\equiv -\frac{\bar{\nu}}{\sigma_{\mathrm{S}}} G^{(1)},  \label{def: A1}  \\
        A_{2} &\equiv \frac{1}{2} \left( \frac{\bar{\nu}^2}{\sigma_{\mathrm{S}}^2} (G^{(1)})^2 - \frac{\bar{\nu}}{\sigma_{\mathrm{S}}} G^{(2)} \right).  \label{def: A2}
    \end{align}
In the above, we retain not only the linear term but also the quadratic term in $\psi_{\mathrm{L}}$.
This is because, as we will see below, the PBH correlation function is proportional to the two-point function of $\psi_{\mathrm{L}}$.
Although it turns out that the quadratic term in Eq.~\eqref{eq: resultant_Pform} does not contribute to the PBH correlation function,
we retain it at this stage for completeness.

Because the inverse function $G$ is used above, it may be useful to replace
$G^{(1)}$ with the quantities directly expressed in terms of the original function $F$
for later purpose.
This can be done by the standard differential formula;
\begin{equation} \label{expression-G1}
G^{(1)}=-\frac{\partial F}{\partial \psi_{\mathrm{L}}}\bigg|_{\rm peak} \bigg/
\frac{\partial F}{\partial \chi_{\mathrm{S}}}\bigg|_{\rm peak},
\end{equation}
where the subscript ``peak'' means that the corresponding quantity
is evaluated at $\chi_{\mathrm{S}}={\bar \nu}\sigma_{\mathrm{S}}, \psi_{\mathrm{L}}=0$.

\subsection{Two-point correlation function for arbitrary local-type non-Gaussianities}
In order to obtain the PBH correlation function, 
we need to compute the two probabilities $P_1$ and $P_2$.
By definition, $P_{1}$ is the probability that a PBH forms in a region centered at a position $\vec{x}$, and $P_{2}$ is the probability that two PBHs simultaneously form in regions centered at $\vec{x}_{1}$ and $\vec{x}_{2}$.
Thus, in terms of $P_{\mathrm{form}}$, by their definitions, they should be given as
\begin{equation}  \label{eq: p1_p2_ito_pform}
    \begin{split}
        P_{1} &= \langle P_{\mathrm{form}}( \psi_{\mathrm{L}}) \rangle,  \\
        P_{2}(\vec{x}_{1}, \vec{x}_{2}) &= \langle P_{\mathrm{form}}(\psi_{\mathrm{L}}(\vec{x}_{1}))P_{\mathrm{form}}(\psi_{\mathrm{L}}(\vec{x}_{2})) \rangle,
    \end{split}
\end{equation}
where $\langle \cdots \rangle$ should be understood as the ensemble average over $\psi_{\mathrm{L}}$.
Substituting Eq.~\eqref{eq: resultant_Pform} and Eq.~\eqref{def: A} into Eq.~\eqref{eq: p1_p2_ito_pform}, we have
\begin{equation} \label{eq: resultant_p1_p2}
\begin{split}
    P_{1} &\simeq \frac{1}{\sqrt{2\pi}\bar{\nu}} e^{-\bar{\nu}^{2}/2} \left(1 +  A_{2} \langle \psi_{\mathrm{L}}^{2} \rangle\right), \\
    P_{2}(\vec{x}_{1}, \vec{x}_{2}) &\simeq \frac{1}{2\pi {\bar \nu}^2}
    e^{-{\bar \nu}^2}\left(1+2A_{2}\langle \psi_{\mathrm{L}}^{2}\rangle+
    A_{1}^{2}\langle \psi_{\mathrm{L}}(\vec{x}_{1}) 
    \psi_{\mathrm{L}}(\vec{x}_{2})\rangle \right) \\
    &\simeq P_{1}^{2}\left(1+A_{1}^{2}\langle \psi_{\mathrm{L}}(\vec{x}_{1}) \psi_{\mathrm{L}}(\vec{x}_{2})\rangle \right).
    \end{split}
\end{equation}
Then, combining Eq.~\eqref{def: pbh_correlator} and \eqref{eq: resultant_p1_p2}, the PBH correlation function in terms of the Gaussian variable 
$\psi_{\mathrm{L}}$ is given as
\begin{equation} \label{eq: resultant_pbh_correlator}
\xi_{\mathrm{PBH}}(\vec{x}_{1}, \vec{x}_{2}) \simeq b^2  
\langle \psi_{\mathrm{L}}(\vec{x}_{1})\psi_{\mathrm{L}}(\vec{x}_{2})\rangle,
\end{equation}
where we have introduced a parameter $b$ as
\begin{equation}
\label{expression-b}
b =\frac{{\bar \nu} \frac{\partial F}{\partial \psi_{\mathrm{L}}}\big|_{\rm peak}}{\sigma_{\mathrm{S}} \frac{\partial F}{\partial \chi_{\mathrm{S}}}\big|_{\rm peak}}.
\end{equation}
This is one of the main results of this work.
This relation relates the PBH correlation function to 
the two-point function of the large-scale field $\psi_{\mathrm{L}}$ 
modulating the small-scale variance of $\zeta_{\mathrm{S}}$.
This formula is valid for arbitrary shapes of LNG 
under the assumption that the large-scale perturbation,
which acts as a source of PBH clustering,
weakly modulates the local PBH abundance. 
The parameter $b$ represents the degree of PBH clustering on large scales 
relative to the underlying large-scale perturbation.
This is analogous to the bias parameter widely used in the structure formation
where the bias parameter acts as a proportionality factor relating galaxy density fluctuations to total matter density fluctuations \cite{Dodelson:2020bqr}.
In this sense, the parameter $b$ may be called the {\it bias parameter} of 
PBH clustering with respect to $\psi_{\mathrm{L}}$.
Note that bias parameter with respect to the large-scale curvature perturbation
was derived in \cite{Tada:2015noa} for the standard LNG: $\zeta=\chi+\frac{3}{5}f_{\rm NL}\chi^2$
and in \cite{Young:2015kda} for the cubic-order LNG: $\zeta=\chi+\frac{6}{5}f_{\rm NL}\chi^2+\frac{9}{25}g_{\rm NL}\chi^3$.

From Eq.~(\ref{expression-b}), it is clear that the initial PBH clustering can arise only
when $\frac{\partial F}{\partial \psi_{\mathrm{L}}}$ is non-zero, namely,
only when the large-scale fluctuation couples to the local(=PBH scale) curvature perturbation.
As a demonstration of our result, in the next subsection, 
we apply Eq.~(\ref{eq: resultant_pbh_correlator}) to concrete examples of the LNG explained in
\ref{sec:benchmark}.

\subsection{Calculations of initial clustering for specific local-type non-Gaussianities} 
\label{sec: example_LNG}
In the previous subsection, we derived the general expression for the PBH correlation function characterizing the initial clustering of PBHs originating from an arbitrary LNG.
This framework enables us to analyze the initial clustering resulting from any inflationary model that predicts LNG, as long as LNG satisfies the assumptions considered in the derivation.
In this section, we explore the physical implications of Eqs.~\eqref{eq: resultant_pbh_correlator} and (\ref{expression-b}) to represent the LNG predicted by concrete inflationary scenarios.

\subsubsection{Power-series non-Gaussianity up to a quadratic order}
The simplest LNG widely studied in the literature is that $\zeta$ is a 
quadratic function of Gaussian field (e.g., \cite{Celoria:2018euj}):
\begin{equation}  \label{def:quad_Lng}
    \zeta(\vec{x}) = \chi\left(\vec{x}\right)+\frac{3}{5}f_{\mathrm{NL}}\chi^{2}(\vec{x}),
\end{equation}
where the so-called non-linearity parameter $f_{\mathrm{NL}}$ 
characterizes the amplitude of the bispectrum in the squeezed limit.
In general, terms higher order in $\chi$ are present in $\zeta$. 
In many cases, they are assumed to be
negligible, which may be justified when the amplitude of $\chi$ (hence $\zeta$) is small. 
This assumption will no longer hold when $\zeta$ is large enough to form a PBH.
Nevertheless, because our current purpose is to demonstrate how to apply our general
formula to specific types of LNG, 
we suppose the form (\ref{def:quad_Lng}) is exact.

Decomposing $\chi$ into short-wavelength mode $\chi_{\mathrm{S}}$ and long-wavelength mode 
$\psi_{\mathrm{L}}$, the function $F$ defined by Eq.~(\ref{def: F}) in the present example becomes
\begin{equation} \label{eq: quad_F}
F=\chi_{\mathrm{S}}+\frac{3}{5} f_{\rm NL}
\left(\chi_{\mathrm{S}}+2\psi_{\mathrm{L}}\right)\chi_{\mathrm{S}}.
\end{equation}
Using this $F$, we can compute quantities appearing in Eq.~(\ref{eq: resultant_pbh_correlator}) as
\begin{equation}
\frac{\partial F}{\partial \psi_{\mathrm{L}}} \Big|_{\rm peak}
=\frac{6}{5} f_{\rm NL}\sigma_{\mathrm{S}} {\bar \nu},~~~~~
\frac{\partial F}{\partial \chi_{\mathrm{S}}} \Big|_{\rm peak}
=1+\frac{6}{5} f_{\rm NL}\sigma_{\mathrm{S}} {\bar \nu}.
\end{equation}
Then, Eq.~(\ref{eq: resultant_pbh_correlator}) in the current example becomes
\begin{equation}
\xi_{\mathrm{PBH}}(\vec{x}_{1}, \vec{x}_{2}) = 
 \left(  
 \frac{ \frac{6}{5} f_{\rm NL} {\bar \nu}^2}
 {1+\frac{6}{5} f_{\rm NL}\sigma_{\mathrm{S}} {\bar \nu} }\right)^{2}
 \langle \psi_{\mathrm{L}}(\vec{x}_{1})\psi_{\mathrm{L}}(\vec{x}_{2})\rangle.
\end{equation}
Despite employing a completely different approach, our formulation shows excellent agreement with the result derived in the Ref.~\cite{Crescimbeni:2025ywm} regarding the non-linear dependence of $b$ on $f_\mathrm{NL}$ in non-perturbative regime.
In the small non-Gaussianity limit $|f_{\rm NL}| \ll 1$,
the above equation leads to $b \approx \frac{6}{5}{\bar \nu}^2f_{\rm NL}$.
In the large $f_{\rm NL}$ limit, $b$ asymptotes to a value $\frac{\bar \nu}{\sigma_{\mathrm{S}}}$
and becomes independent of $f_{\rm NL}$.

\subsubsection{Logarithmic local-type non-Gaussianities}
The non-linear curvature perturbation is predicted to have logarithmic
dependence on the Gaussian variable
in several inflation scenarios described in \ref{sec:benchmark}, such as 
the USR scenario, the curvaton model \cite{Ferrante:2023bgz}, 
and modulated reheating scenario \cite{Suyama:2007bg, You:2024hit}.
To deal with these cases in a unified manner, 
let us assume the curvature perturbation is given as 
\begin{equation}
F(\chi_{\mathrm{S}},\psi_{\mathrm{L}}) = \alpha \ln\left[A(\chi_{\mathrm{S}}+\psi_{\mathrm{L}})\right]-
\alpha \ln\left[A(\psi_{\mathrm{L}})\right],
\end{equation}
where $\alpha$ is constant and $A(x)$, satisfying $A(0)=1$, is a polynomial in $x$ of degree at most two.
For this $F$, we have
\begin{equation}
\frac{\partial F}{\partial \psi_{\mathrm{L}}} \Big|_{\rm peak}
=\alpha\left(\frac{ A'(\sigma_{\mathrm{S}} {\bar \nu})}{A(\sigma_{\mathrm{S}} {\bar \nu})}
-\frac{ A'(0)}{A(0)}\right),~~~~~
\frac{\partial F}{\partial \chi_{\mathrm{S}}} \Big|_{\rm peak}
=\alpha \frac{ A'(\sigma_{\mathrm{S}} {\bar \nu})}{A(\sigma_{\mathrm{S}} {\bar \nu})}.
\end{equation}
Then, Eq.~(\ref{eq: resultant_pbh_correlator}) becomes
\begin{equation}
\label{pbh-xi-log}
\xi_{\mathrm{PBH}}(\vec{x}_{1}, \vec{x}_{2}) = 
{\left(  
\frac{\bar \nu}{\sigma_{\mathrm{S}}} 
\left( 1-\frac{ A(\sigma_{\mathrm{S}} {\bar \nu}) }{A(0)} \frac{ A'(0) }{A'(\sigma_{\mathrm{S}} {\bar \nu})} \right)
\right)}^2
\langle \psi_{\mathrm{L}}(\vec{x}_{1})\psi_{\mathrm{L}}(\vec{x}_{2})\rangle.
\end{equation}
In what follows, we apply this result to a couple of examples. \\

\underline{Ultra slow-roll scenario} 

The first example is the USR scenario explained as a second example in \ref{sec:benchmark}.
Because our purpose in the present section is to demonstrate application of our formula to concrete
examples, we assume the limiting case where the USR phase ends instantly. 
In this case, the resultant small-scale curvature $\zeta_{\mathrm{S}}$ produced during the USR phase 
is given by Eq.~(\ref{USR-zetas}).
This corresponds to a case 
\begin{equation}
    \alpha=-\frac{1}{3},~~~~~A(x) =1-3x.
\end{equation}
Then, Eq.~(\ref{pbh-xi-log}) becomes
\begin{equation}  \label{pbh-xi-usl}
\xi_{\mathrm{PBH}}(\vec{x}_{1}, \vec{x}_{2}) = 
9{\bar \nu}^4
\langle \psi_{\mathrm{L}}(\vec{x}_{1})\psi_{\mathrm{L}}(\vec{x}_{2})\rangle.
\end{equation}
\\

\newpage

\underline{Curvaton scenario} 

As in \ref{sec:benchmark}, we consider a limiting case where the curvaton
dominates the universe before the curvaton decays into radiation.
In this case, $\alpha$ and $A(x)$ are given by
\begin{equation}
\alpha = \frac{2}{3},~~~~~A(x) = 1+\frac{3}{2}x.
\end{equation}
Then, Eq.~(\ref{pbh-xi-log}) becomes
\begin{equation}  \label{pbh-xi-curv}
\xi_{\mathrm{PBH}}(\vec{x}_{1}, \vec{x}_{2}) = 
\frac{9}{4}{\bar \nu}^4
\langle \psi_{\mathrm{L}}(\vec{x}_{1})\psi_{\mathrm{L}}(\vec{x}_{2})\rangle.
\end{equation}
where $\psi_{\mathrm{L}}$ is related to the curvaton field $\varphi$ by
$\psi_{\mathrm{L}}=(2/3)\cdot(\delta \varphi_{\mathrm{L}}/{\bar \varphi})$.
Comparing Eq.~\eqref{pbh-xi-usl} and Eq.~\eqref{pbh-xi-curv}, one might expect that the PBH initial clustering occurs more strongly in the USR scenario than in the curvaton scenario.
However, this is not necessarily the case because the relative strength depends on the choice of
a constant factor in the definition of $\psi_{\mathrm{L}}$.
We emphasize that what is physically meaningful is the product of the PBH bias parameter and the correlation function of $\psi_{\mathrm{L}}$.
\\

\underline{Modulated reheating scenario} 

The last example is the modulated reheating scenario.
If inflaton $\phi$ has a decay channel whose corresponding interaction depends on the expectation
value of the light field $\varphi$, the corresponding decay rate has spatial fluctuations tracing the fluctuations of $\varphi$.
For instance, the interaction term given by $\frac{\varphi}{\Lambda} \phi {\bar f}f$,
where $\Lambda$ is constant of mass dimension 1 and $f$ is fermionic field,
leads to a decay rate given by $\frac{m}{8\pi}\frac{\varphi^2}{\Lambda^2}$ ($m$ is the mass of inflaton).
Generally, the total decay rate $\Gamma$ consists of a component independent of the value 
of the other field and a remaining part that depends on it.
Then, the total decay rate takes the following form
\begin{equation}  \label{modulated_decay_rate}
\Gamma =\Gamma_0+\Gamma_1 \varphi^2,
\end{equation}
where $\Gamma_0$ and $\Gamma_1$ are constants.
We adopt this $\Gamma$ in the following calculations.

Using that the curvature perturbation in the modulated reheating scenario is given by Eq.~(\ref{zetas-modulated}),
we find the corresponding $\alpha$ and $A(x)$ are given by
\begin{equation}
\alpha= -\frac{1}{6},~~~~~A(x)=1-6 \left(2x-\frac{6}{\beta}x^2\right),
~~~~~\beta \equiv \frac{\Gamma_1 {\bar \varphi}^2}{\Gamma_0+\Gamma_1 {\bar \varphi}^2}.
\end{equation}
Then, Eq.~(\ref{pbh-xi-log}) becomes
\begin{equation}
\xi_{\mathrm{PBH}}(\vec{x}_{1}, \vec{x}_{2}) = 36
{\bar \nu}^4 {\left( \frac{1-\beta (2+\sigma_{\mathrm{S}} {\bar \nu})}{\beta(1+\sigma_{\mathrm{S}} {\bar \nu})} \right)}^2
\langle \psi_{\mathrm{L}}(\vec{x}_{1})\psi_{\mathrm{L}}(\vec{x}_{2})\rangle.
\end{equation}
where $\psi_{\mathrm{L}}$ is related to the light field $\varphi$ by
$\psi_{\mathrm{L}}=-(\beta/6)\cdot(\delta \varphi_{\mathrm{L}}/{\bar \varphi})$
\footnote{Since $\Gamma'/\Gamma = \beta/\bar{\varphi}$ with Eq.~\eqref{modulated_decay_rate}, this is the same definition as $\psi_\mathrm{L}$ in Eq.~\eqref{zetas-modulated}}.

\subsection{Connection to the primordial trispectrum}
As we have discussed in \ref{subsec:intuitive}, the right-hand side of Eq.~\eqref{eq: resultant_pbh_correlator} can be connected to the trispectrum
of the primordial curvature perturbation in the collapsed limit.
To see this, we first note that the large-scale correlation
of the variance of $\zeta_{\mathrm{S}}$ is given by (see the derivation in Appendix~\ref{four_point_correlation})
\begin{equation}  \label{eq: relation_bw_short_and_Long_modes}
\left\langle\zeta_{\mathrm{S}}^{2}\left(\vec{x}_{1}\right)\zeta_{\mathrm{S}}^{2}\left(\vec{x}_{2}\right)\right\rangle_{\mathrm{c}}=
4\tau_{\rm NL} {\langle \zeta_{\mathrm{S}}^2 \rangle}^2
\langle \zeta_\mathrm{L} (\vec{x_1}) \zeta_\mathrm{L} (\vec{x_2}) \rangle,
\end{equation}
where the subscript c represents the connected part and 
$\tau_{\mathrm{NL}}$ is the non-linearity parameter characterizing the
magnitude of the trispectrum in the collapsed limit (see 
Eq.~(\ref{def-taunl}) for its definition).
Meanwhile, using Eq.~(\ref{def: F}) we can explicitly compute 
$\langle\zeta_{\mathrm{S}}^{2}(\vec{x}_{1})\zeta_{\mathrm{S}}^{2}
(\vec{x}_{2})\rangle_{\mathrm{c}}$.
To see this, we expand Eq.~(\ref{def: F}) in $\psi_{\mathrm{L}}$ as
\begin{equation}
    \zeta_{\mathrm{S}}(\vec{x}) = \sum_{n=0}^\infty \frac{1}{n!}
    F^{(n)}(\chi_{\mathrm{S}}(\vec{x})) {( \psi_{\mathrm{L}}(\vec{x}) )}^n,
\end{equation}
where
\begin{equation} \label{def: fn}
    F^{(n)}(\chi_{\mathrm{S}}) \equiv \frac{\partial^{n}F}{\partial\psi_\mathrm{L}^{n}}\Bigg|_{\chi_{\mathrm{S}}=\chi_{\mathrm{S}}, \psi_{\mathrm{L}}=0}.
\end{equation}
Then, using the fact that $\chi_{\mathrm{S}}$ and $\psi_{\mathrm{L}}$ are uncorrelated,
to the leading order in $\psi_{\mathrm{L}}$, we have 
\begin{equation} \label{eq: resultant_four_point_ito_chi}
    \left\langle\zeta_{\mathrm{S}}^{2}\left(\vec{x}_{1}\right)\zeta_{\mathrm{S}}^{2}\left(\vec{x}_{2}\right)\right\rangle_{\mathrm{c}} 
    \simeq 
    4 \langle F^{(0)}(\chi_{\mathrm{S}}) F^{(1)}(\chi_{\mathrm{S}})\rangle^2 \left\langle\psi_{\mathrm{L}}(\vec{x}_{1})\psi_{\mathrm{L}}(\vec{x}_{2})\right\rangle.  
\end{equation}
This clearly shows that the presence of the coupling between $\psi_{\mathrm{L}}$ and the small-scale curvature perturbation (i.e., $F^{(1)}$) leads to the long-range correlation of the local variance
of $\zeta_{\mathrm{S}}$.
From Eqs.~(\ref{eq: relation_bw_short_and_Long_modes}) and (\ref{eq: resultant_four_point_ito_chi}),
we can relate $\langle\psi_{\mathrm{L}}(\vec{x}_{1})\psi_{\mathrm{L}}(\vec{x}_{2})\rangle$ to 
$\langle\zeta_L(\vec{x}_{1})\zeta_L(\vec{x}_{2})\rangle$ and use it to rewrite Eq.~(\ref{eq: resultant_pbh_correlator}) in terms of 
$\langle\zeta_L(\vec{x}_{1})\zeta_L(\vec{x}_{2})\rangle$. 
Substituting Eq.~(\ref{expression-G1}) as well, 
we finally obtain
\begin{equation} \label{eq: generic_pbh_correlator}
    \xi_{\mathrm{PBH}}(\vec{x}_{1}, \vec{x}_{2}) \simeq b_\zeta^2
    \langle\zeta_\mathrm{L}(\vec{x}_{1})\zeta_\mathrm{L}(\vec{x}_{2})\rangle, 
\end{equation} 
where the bias parameter $b_\zeta$ defined for the case where the underlying field is curvature perturbation
is given by
\begin{equation} \label{PBH-bias-parameter}
   b_\zeta= \tau^{1/2}_{\rm NL} \frac{{\bar \nu} \frac{\partial F}{\partial \psi_{\mathrm{L}}}
   \big|_{\rm peak}
   \langle F^{(0)}(\chi_{\mathrm{S}}) F^{(0)}(\chi_{\mathrm{S}}) \rangle }{\sigma_{\mathrm{S}} \frac{\partial F}{\partial \chi_{\mathrm{S}}} 
   \big|_{\rm peak}
   \langle F^{(0)}(\chi_{\mathrm{S}}) F^{(1)}(\chi_{\mathrm{S}})\rangle}.
\end{equation}
As desired, we have been able to express the bias parameter in terms
of a quantity characterizing the collapsed limit of the trispectrum which represents the
large-scale modulation of the local(=PBH scale) variance of the curvature perturbation.

In the previous study \cite{Suyama:2019cst},  
$b_\zeta$ in the limit of weak non-Gaussianity given by Eq.~(\ref{def:quad_Lng}) was obtained
based on the functional integration approach.
To check consistency with this result, 
we evaluate Eq.~(\ref{eq: quad_F}) up to the leading order in $f_{\rm NL}$.
Then, we have $\langle F^{(0)}(\chi_{\mathrm{S}}) F^{(0)}(\chi_{\mathrm{S}}) \rangle=\sigma_{\mathrm{S}}^2$,
$\langle F^{(0)}(\chi_{\mathrm{S}}) F^{(1)}(\chi_{\mathrm{S}})\rangle=\frac{6}{5}f_{\rm NL} \sigma_{\mathrm{S}}^2$, $\frac{\partial F}{\partial \psi_{\mathrm{L}}}\big|_{\rm peak}=\frac{6}{5}f_{\rm NL} \sigma_{\mathrm{S}} {\bar \nu}$, and $\frac{\partial F}{\partial \chi_{\mathrm{S}}} 
   \big|_{\rm peak}=1$.
Substituting these into Eq.~(\ref{PBH-bias-parameter}) yields $b_\zeta=\tau_{\rm NL}^{1/2}{\bar \nu}^2$,
which exactly reproduces the result in \cite{Suyama:2019cst}.
Notice that $\tau_{\rm NL}$ is related to $f_{\rm NL}$ as $\tau_{\rm NL}=\frac{36}{25}f_{\rm NL}^2$ in the present case \cite{Byrnes:2006vq}.
Thus, $b_\zeta$ can be written as $b_\zeta=\frac{6}{5} f_{\rm NL}{\bar \nu}^2$.

\subsection{Extension to broad power spectrum}
\label{extension-broad}
Up to this stage, we have assumed that the curvature perturbation is largely enhanced
only at around a specific scale.
In this subsection, we argue that at least at the formal level,
it is possible to generalize the PBH correlation function to the case of the broad 
power spectrum in which PBHs may be abundantly produced over a wide range of PBH mass.
In what follows, we assume such enhanced curvature perturbation is correlated with 
large-scale fluctuation which weakly modulates the small-scale curvature perturbations.
(Large scale is meant to be scale larger than scales PBH formation occurs and small scales refer
to those where PBH formation occurs.)

We again suppose the curvature perturbation takes a local form
given by Eq.~(\ref{def: general_LNG}), with only $\chi$ and $\psi$ being the argument of $\zeta$.
As in the previous case, $\chi$ is enhanced at small scales and $\psi$ has only large-scale component
(in what follows, we change the notation to $\psi_{\mathrm{L}}$).
When the perturbations are enhanced over a wide range of scales,
PBH formation happens at different times with different masses.
PBH with a specific mass is formed when the curvature perturbation with the corresponding scale $S$ reenters
the Hubble horizon.
Given that curvature perturbations at different scales do not affect 
whether the PBH at the scale $S$ is formed or not,
important quantity is the curvature perturbation at a specific scale $S$ which 
is obtained by extracting the corresponding modes from the original $\zeta$ as
\begin{equation}  \label{def: local_zeta}
    \zeta_{\mathrm{S}}(\vec{x}) = \int{d^{3}y}~W\left(S,\left|\vec{x}-\vec{y}\right|\right) 
    f (\chi(\vec{y}),\psi_{\mathrm{L}} (\vec{y}) ),
\end{equation}
where $W(S,r)$ is a function which retain only components varying over the scale $S$ \cite{Suyama:2019cst}.
Assuming $\psi_{\mathrm{L}}$ can be treated as linear perturbation, we expand Eq.~(\ref{def: local_zeta}) as
\begin{align}
\label{broad-zetaS}
\zeta_{\mathrm{S}}(\vec{x})&=\int d^3y~W(|\vec{y}-\vec{x}|,S)
\left( f (\chi (\vec{y}),0) +\frac{\partial f}{\partial \psi_{\mathrm{L}}}\bigg|_{\psi_{\mathrm{L}}=0} \psi_{\mathrm{L}} (\vec{y}) \right) \nonumber \\
&=F^{(0)}[\chi,S,\vec{x}]+F^{(1)}[\chi,S,\vec{x}] \psi_{\mathrm{L}} (\vec{x}),
\end{align}
where the coefficients $F^{(0)}$ and $F^{(1)}$ are easily read by comparing the first line
with the second line.
In arriving at the second line, we have used that $\psi_{\mathrm{L}}$ varies little on the scale $S$
which gives the effective volume of the integral over $\vec{y}$.

When the overdense region with its size $S$ has $\zeta_{\mathrm{S}}$ exceeding the threshold, 
that region collapses to a PBH with mass $m$.
For simplicity, we ignore the effect of the critical phenomena and adopt a picture 
that the PBH mass is equal
to the horizon mass at the time of PBH formation.
Using Eq.~(\ref{broad-zetaS}), the probability that a PBH with mass $m$ forms at $\vec{x}$
is given by
\begin{equation}
P_{\rm form} (m,\psi_{\mathrm{L}}(\vec{x}))=\int_{\zeta_{\mathrm{S, th}}} d\zeta_{\mathrm{S}} 
\int {\cal D}\chi~\delta (\zeta_{\mathrm{S}}-F^{(0)}[\chi,S,\vec{x}]-F^{(1)}[\chi,S,\vec{x}] \psi_{\mathrm{L}} (\vec{x})
) ~{\rm PDF}[\chi],
\end{equation}
where ${\rm PDF}[\chi]$, which is the functional of $\chi$, is the PDF of $\chi$.
Expanding the Dirac's delta function up to first order in $\psi_{\mathrm{L}}$, we obtain
\begin{equation}
\label{broad:Pform}
P_{\rm form} (m,\psi_{\mathrm{L}}(\vec{x}))=B_0(S)+B_1(S) \psi_{\mathrm{L}} (\vec{x}),
\end{equation}
where $B_0$ and $B_1$ are given by
\begin{equation}
B_0 (S) =\int_{\zeta_{\mathrm{S, th}}} d\zeta_{\mathrm{S}} ~{\rm PDF}(\zeta_{\mathrm{S}}),~~~
B_1 (S)=\int {\cal D}\chi~\delta (\zeta_{\mathrm{S, th}}-F^{(0)}[\chi,S]) F^{(1)}[\chi,S]~{\rm PDF}[\chi],
\end{equation}
and ${\rm PDF}(\zeta_{\mathrm{S}})=\int {\cal D}\chi~\delta(\zeta_{\mathrm{S}}-F^{(0)}[\chi,S])~{\rm PDF}[\chi]$
is the PDF of $\zeta_{\rm S}$.

Because enhanced perturbations at different scales reenter the Hubble horizon at different times,
PBHs with different masses are formed.
Let us pick up two different scales $S_1$ and $S_2$ in the range where PBHs are produced and
denote by $m_1$ and $m_2$ the PBH mass corresponding to the scale $S_1$ and $S_2$, respectively.
The PBH correlation function measuring the probability excess of finding
a PBH with $m_1$ at $\vec{x_1}$ and another PBH with $m_2$ at $\vec{x_2}$ can be written as
\begin{equation}
\xi_{\rm PBH}(m_1,\vec{x_1};m_2,\vec{x_2})=
\frac{P_2(m_1,\vec{x_1};m_2,\vec{x_2})}{P_1 (m_1,\vec{x_1}) P_1 (m_2, \vec{x_2})}-1,
\end{equation}
where the meaning of $P_1$ and $P_2$ is the same as Eq.~(\ref{def: pbh_correlator}).
Using Eq.~(\ref{broad:Pform}), we obtain
\begin{equation}
\xi_{\rm PBH}(m_1,\vec{x_1};m_2,\vec{x_2})=\frac{B_1 (S_1) B_1 (S_2)}{B_0 (S_1) B_0 (S_2)} \langle \psi_{\mathrm{L}} (\vec{x_1})
\psi_{\mathrm{L}} (\vec{x_2})\rangle.
\end{equation}
This may be considered as a formal generalization of Eq.~(\ref{eq: resultant_pbh_correlator}) to
correlation function between different PBH masses.
More explicit form of $B_1$ is available only after the form of $\zeta$ as a function of $\chi_{\mathrm{S}}$ and $\psi_{\mathrm{L}}$ is explicitly specified. 
However, computations of $B_1 $, which involves functional integral, will be complicated in general.

\section{Conclusion}  \label{sec: conlusion}
In this work, we have investigated the initial spatial clustering of primordial black holes 
(PBHs) induced by an arbitrary local-type non-Gaussianity (LNG) 
of the primordial curvature perturbations. 
Previous studies on PBH initial clustering have primarily restricted their analyses to LNG in the limited, extreme regimes.
However, the LNG predicted by several benchmark inflationary scenarios 
yielding substantial PBH formation does not necessarily fall into such regimes.
While this motivated recent work to address arbitrary LNG~\cite{Crescimbeni:2025ywm},
our study successfully establishes a general analytical formulation for the PBH two-point 
correlation function that is valid for any arbitrary LNG
via a distinct approach to that of Ref.~\cite{Crescimbeni:2025ywm}.
Our derivation relies on the assumptions that PBH formation occurs in the high-peak limit, 
where the threshold for gravitational collapse is very high, 
and that the large-scale mode only weakly modulates the local variance of 
the small-scale curvature perturbations. 
Under this framework, we have demonstrated that the PBH two-point correlation function 
$\xi_{PBH}$ is proportional to the correlation function of the underlying 
large-scale Gaussian field $\psi_L$, scaled by a bias parameter $b$. 
We explicitly showed that the source of this initial clustering is governed by the correlation 
of the local variance of small-scale perturbations, 
which is directly connected to the primordial trispectrum in the collapsed limit parameterized by $\tau_{NL}$. 
By utilizing our general formula, we successfully quantified the initial clustering for several widely studied inflationary scenarios predicting non-trivial LNGs, 
specifically the ultra-slow-roll inflation, curvaton, and the modulated reheating scenarios. 
Furthermore, we formally generalized our framework to incorporate broad power spectra
and derived the analytical expression of spatial correlations between PBHs of different masses.  

The most significant advantage of our formulation is its applicability to arbitrary, 
non-perturbative LNGs. 
This provides a robust theoretical foundation for incorporating the effects of strong LNGs 
into the analysis of PBH spatial distributions. 
In future work, this comprehensive framework can be utilized to accurately assess 
clustering-induced modifications to various cosmological constraints, 
such as the binary black hole merger rates in the early and late universe, 
microlensing event rates, and the epoch of first structure formation.

\section*{Acknowledgement}
This work was supported by JSPS KAKENHI Grant Number JP23K03411 (TS).
The work of A.K. was supported by the Japan Society for the Promotion of Science (JSPS) as part of the JSPS Postdoctoral Program (Grant Number: 25KF0107).

\appendix

\section{Derivation of Eq.~(\ref{eq: relation_bw_short_and_Long_modes})}  
\label{four_point_correlation}
Let $\zeta_{\mathrm{S}}$ be the small-scale curvature perturbation relevant to PBH formation. 
Then, the connected part of the four-point function $\langle\zeta_{\mathrm{S}}^{2}(\vec{x}_{1})\zeta_{\mathrm{S}}^{2}(\vec{x}_{2})\rangle$ in terms of the Fourier components becomes
\footnote{
The Fourier transformation ${\tilde f}_{\vec{k}}$ of the function $f(\vec{x})$ is defined by
\begin{align}
    \tilde{f}_{\vec{k}} \equiv \int{d^{3}x}~e^{-i\vec{k}\cdot\vec{x}}f(\vec{x}),\qquad f(\vec{x}) \equiv \int\frac{d^{3}k}{\left(2\pi\right)^{3}}e^{i\vec{k}\cdot\vec{x}}\tilde{f}_{\vec{k}}.
\end{align}
} 
\begin{align}
\label{4point-zetas}
    \left\langle\zeta_{\mathrm{S}}^{2}\left(\vec{x}_{1}\right)\zeta_{\mathrm{S}}^{2}\left(\vec{x}_{2}\right)\right\rangle_{\mathrm{c}} &= 
    \int\frac{d^{3}p_{1}}{\left(2\pi\right)^{3}}
    \frac{d^{3}q_1}{\left(2\pi\right)^{3}}
    \frac{d^{3}p_2}{\left(2\pi\right)^{3}}
 e^{-i (\vec{p_1}+\vec{q_1})\cdot (\vec{x_1}-\vec{x_2})}
    T_\zeta(\vec{p_1},\vec{q_1},\vec{p_2},-\vec{p_1}-\vec{q_1}-\vec{p_2}),
\end{align}
where we have used the definition of the primordial trispectrum $T_\zeta$ given by
\be
\left\langle\tilde{\zeta}_{\vec{p_1}}\tilde{\zeta}_{\vec{p_2}}
    \tilde{\zeta}_{\vec{p_3}}\tilde{\zeta}_{\vec{p_4}}\right\rangle_{\mathrm{c}}
    ={(2\pi)}^3 \delta (\vec{p_1}+\vec{p_2}+\vec{p_3}+\vec{p_4})~
    T_\zeta (\vec{p_1},\vec{p_2},\vec{p_3},\vec{p_4}).
\ee
Because $\zeta_{\mathrm{S}}$, in Fourier space, does not vanish only around the scale
of PBH formation,
all the momenta appearing as arguments of the trispectum have momenta around the scale of PBH formation.
With this in mind, let us take $|\vec{x_1}-\vec{x_2}|$ to be much larger than the PBH scale.
Then, the phase of the exponential function 
$e^{-i (\vec{p_1}+\vec{q_1})\cdot (\vec{x_1}-\vec{x_2})}$
becomes very large except for quadrangle configurations with 
$|\vec{p_1}+\vec{q_1}| \lesssim 1/|\vec{x_1}-\vec{x_2}|$.
Thus, in the large distance limit,
the trispectrum in the collapsed limit where
$|\vec{p_1}+\vec{p_2}|\ll p_i$ ($i=1,\cdots,4$)
dominantly contributes to the integral.
It is standard to parametrize the magnitude of the trispectrum in that limit as (e.g., \cite{Smith:2011if})
\be
\label{def-taunl}
\tau_{\rm NL}=\frac{1}{4} \lim_{|\vec{p_1}+\vec{p_2}|\to 0}
\frac{T_\zeta (\vec{p_1},\vec{p_2},\vec{p_3},\vec{p_4})}{P_\zeta (p_1)
P_\zeta (p_3) P_\zeta (|\vec{p_1}+\vec{p_2}|)}.
\ee
Using this parameter, Eq.~(\ref{4point-zetas}) in the large distance limit becomes
\be
\left\langle\zeta_{\mathrm{S}}^{2}\left(\vec{x}_{1}\right)\zeta_{\mathrm{S}}^{2}\left(\vec{x}_{2}\right)\right\rangle_{\mathrm{c}}=
4\tau_{\rm NL} {\langle \zeta_{\mathrm{S}}^2 \rangle}^2
\langle \zeta_\mathrm{L} (\vec{x_1}) \zeta_\mathrm{L} (\vec{x_2}) \rangle,
\ee
where
\be
\langle \zeta_{\mathrm{S}}^2 \rangle =\int \frac{d^3q}{{(2\pi)}^3} P_\zeta (q)
\ee
is the variance of $\zeta_{\mathrm{S}}$ and 
\be
\langle \zeta_\mathrm{L} (\vec{x_1}) \zeta_\mathrm{L} (\vec{x_2}) \rangle
=\int \frac{d^3q}{{(2\pi)}^3} e^{-i \vec{p}\cdot (\vec{x_1}-\vec{x_2})}
P_\zeta (p)
\ee
is the two-point function of $\zeta$ at large distance.

\bibliography{ref}
\end{document}